Aligned Major Axes in a Planetary System without Tidal Evolution:

The 61 Virginis example

by Richard Greenberg and Christa Van Laerhoven

Lunar and Planetary Laboratory
University of Arizona
1629 East University Boulevard
Tucson, AZ 85721-0092

Manuscript # MN-10-1863-MJ (Revised)

Corresponding author:
Richard Greenberg
greenberg@lpl.arizona.edu

Aligned Major Axes in a Planetary System without Tidal Evolution:

The 61 Virginis example

**Abstract**

Tidal damping of one of the orbits in a planetary system can lead to aligned major-axes (the so-called "fixed-point" condition), but currently aligned major axes do not necessarily imply such a history. An example is the nominal orbital solution for the 61 Virginis system where two orbits librate about alignment, but evaluation of the eigenmodes of the secular theory shows it could not be the result of tidal damping but rather of initial conditions. Nevertheless, the amplitudes of the eigenmodes suggest that this system may have undergone some degree of tidal damping.

Key Words: Celestial mechanics, Planets and satellites: dynamical evolution and stability

**1. Introduction**

The accelerating discoveries of multi-planet extra-solar systems are placing revealing constraints on the origin and evolution of planetary systems. The architecture of each system results from the process of planet formation; from dynamical effects of the nebula from which they formed (e.g. Lin et al. 1996); from close approaches among the planets as they settled toward a relatively stable configuration (e.g. Rasio and Ford 1996); from tidal effects in the many systems with past or present close-in planets (e.g. Trilling 2000, Jackson et al. 2008), and from on-going gravitational interactions among the planets. Orbital resonances enhance these mutual perturbations in many cases (e.g. Lee and Peale 2002), but even without resonances, secular interactions are significant and the character of those interactions can provide important insights into the history of a system.

While only a close-in planet is directly affected by tides, secular interactions can distribute the angular momentum causing long-term changes in the orbits of all the planets. Thus the current, coupled dynamical state of the entire system must be considered as a basis for any inferences about tidal evolution over the history of the system. For example, if the orbital major axes of the planets in a system are aligned (or librating about alignment) it may be the result of a process (e.g. tides) that tended to damp the eccentricity of one of the planets (Wu and Goldreich 2002, Zhang and Hamilton 2003). Consequently, aligned major axes are often interpreted as a signature of tidal evolution. However, we demonstrate here that this is not necessarily the case, unless certain conditions of the current system are met.

The nominal parameters of the 61 Virginis planetary system (Vogt et al. 2010) provide an instructive example. This system is also particularly noteworthy because the masses of the planets are comparable to those in our solar system (Table 1). The inner planet has a mass comparable to that of the Earth, while the two farther out planets have masses comparable to Neptune's. All three planets are quite close to their star by solar system standards, strengthening their mutual perturbations, and the inner planet is close enough that tidal evolution has likely been important. The observational best-fit keplerian orbital elements are shown in Table 1. Vogt et al. found that an adequate fit to the radial-velocity data could be made assuming perfectly circular orbits as well. However, for our purposes, we only need an illustrative example, so we consider the behavior of the system assuming the best-fit elements.

Even if accurate, current orbital elements obscure the dynamical character of any system, because the secular interactions periodically change the values. For the nominal solution for 61 Virginis, elements vary considerably over timescales of $\sim 10^4$ years (e.g. Vogt et al., Fig. 8). Eccentricities are not constants of integration in multi-planet systems. The character of the system can be more meaningfully represented by the constants of integration from secular theory.

Considerable insight can come from interpreting planetary systems in terms of the standard second-order secular theory (e.g. Wu and Goldreich 2002, Chiang and Murray 2002, Zhang and Hamilton 2003, Adams and Laughlin 2006, Barnes and Greenberg 2006), even though the actual behavior may differ to some degree due to substantial eccentricities (as discussed for example by Ford et al. (2000) and Veras and Armitage (2007)).  Applying this classical secular theory below, and including eccentricity damping, we review the basis for the association of aligned major axes with past tidal evolution; we confirm Vogt et al.'s finding that in the nominal 61 Virginis system the major axes of the two outer planets librate about alignment; and we show why in this case that alignment is not indicative of tidal damping of the inner planet's orbit, contrary to the interpretation by Vogt et al. As we show below, the libration about orbital alignment more likely reflects initial conditions of the eccentricities, abetted by a modest propensity for such alignment due to the particular masses and semi-major axis values in the system.

This study shows that, while eccentricity damping can eventually lead to a long-lived condition of aligned major axes, a system with aligned major axes is not necessarily indicative of such a history.  This work also exemplifies the value of classical secular theory in classifying and interpreting the dynamical characteristics of multi-planet systems.

## 2.  Current Secular Behavior

The essential points of classical secular theory are summarized in Appendix A. The eccentricity vector (defined by the $e$ value and the perihelion longitude $\varpi$) for each planet in an $N$-planet system is a sum of $N$ vector components (Fig. 1).  Each component (a) corresponds to one of $N$ eigenmodes; (b) has a fixed magnitude; and (c) rotates at a rate given by the eigenfrequency.

Applying Appendix A to the nominal (i.e. best-fit) semi-major axes and masses in the 61 Virginis system (Table 1), yields the eigenfrequencies $g_m$ (where the eigenmode is identified by the subscript $m = 1, 2,$ or 3) and the corresponding normalized eigenvectors $V_{mp}$ (Table 2).  For this demonstration, we take the planetary masses to be equal to the observational determined values of $M \sin i$. Then, the current $e$ and $\varpi$ values yield the constants of integration:  the magnitudes of the eigenvectors $E_m$ and the phases $\delta_m$. (Table 2).  From those results, Fig. 2 shows the current eccentricity vectors for planets 1, 2, and 3 (61 Vir b, c, and d, respectively), each as a vector sum of its three eigenmode components ($E_m V_{mp}$). The secular solution (Eqs. A6) requires that the vectors for all the planets for any given eigenmode (i.e. any given value of $m$) are aligned (or anti-aligned) and rotate at the same rate (the eigenfrequency) as one another. The alignment for each eigenmode is evident in Fig. 2.

In any system in which the amplitudes of all eigenvectors except one are zero (sometimes called a "fixed-point" configuration), the major axes of the planets would be locked into alignment, because the circulation of their $e$ vectors would all be in phase. Also, in that case, the eccentricity values would be constant.

In the case of 61 Virginis, the single eigenmode ($m = 3$) dominates for the outer two planets. (This result follows assuming the best-fit solution, as shown in Table 2, and indeed holds true over most of the range of uncertainty in the orbital solution, as discussed in Section 4.3 below.) Thus the orientation of the major axes of those planets are locked to (albeit librating about) alignment. Vogt et al. suggested that this behavior might indicate a "fixed point" configuration, which they interpreted as a sign that tidal evolution has occurred, because tides tend to leave a single long-lived eigenmode after relatively quickly damping out the others.

However inspection of Fig. 2 shows that none of the eigenmode components has been damped down to near zero. Although the $m=3$ components for the two outer planets are great enough compared with the other components to keep the major axes of those two planets from circulating, the amplitudes for the $m=1$ and 2 eigenmodes are also substantial. For example, the $m=2$ component of the eccentricity of planet #1 (0.219) is practically as large as the $m=3$ part for planet #2 (0.224). Thus this is not a fixed-point configuration and, as discussed in the next section, this system shows no evidence of eccentricity damping.

## 3. Effects of eccentricity damping

To understand why tidal damping has been associated with aligned major axes, suppose the inner planet has been affected by an eccentricity-damping process described by $de_1/dt = -Fe_1$.

Tides are an effective process for damping eccentricities in this way. (Here we ignore accompanying changes in the semi-major axis.) As reviewed in Appendix B, the equations that govern secular behavior remain linear when this effect is included, but now the solution includes an imaginary component in each eigenfrequency, which translates into an exponential damping of each eigenmode. In physical terms, the secular interactions exchange angular momentum, but not energy, in such a way that the eccentricity damping is shared among them.

Numerically evaluating the solution formula (Eq. B5) using the best-fit orbital elements, we find that the eigenvectors for modes $m = 1$, 2, and 3 damp at exponential rates $0.65F$, $0.35F$, and $0.001F$, respectively. In other words, the first two eigenmodes damp about half as fast as the inner planet would have alone. The third eigenmode would require a timescale three orders of magnitude longer to damp, and is thus the long-lived, so-called "fixed-point" solution. It indeed would be characterized by alignment of all three major axes. However, as we have seen in the previous section, for 61 Virginis, none of the eigenmodes appears to have damped significantly, and thus there is no evidence for tidal evolution in this system in the currently available data.

One might intuitively expect the eigenmode that dominates the behavior of the inner planet to be the one that damps fastest when the eccentricity of this planet is damped. However, that relationship does not always hold. In this case, Fig. 2 shows that mode 2 is dominant for 61 Vir b, while the above result shows that mode 1 damps almost twice as fast.

It is also instructive to consider what would happen if some hypothetical process damped the eccentricity of one of the other planets by some mechanism. Again, the damping would be shared among all the planets. For example, although tides are unlikely to have an effect on the second planet in this system because of its distance from the star, if some process did tend to damp it with an exponential rate $F_2$, an analysis similar to that in Appendix B shows that eigenmodes 1, 2 and 3 would damp at rates $0.29F_2$, $0.49F_2$, and $0.22F_2$, respectively. In that hypothetical case, all three eigenmodes would damp over similar timescales. The "fixed-point" condition (with only the longest lived eigenmode remaining) would only occur after all the eccentricities had damped significantly, and it would not survive much longer than the damping timescale for the other eigenmodes. In other words, in general the so-called "fixed-point" condition may not necessarily be fixed for long.

## 4. Discussion

### 4.1 Cause of the alignment

If not due to tidal damping, an alignment of major axes, as in the nominal 61 Virginis case, can be simply a result of the initial conditions of the eccentricity vectors (which define the magnitudes and phases of the eigenvectors) within the context of the basic architecture of the system (the planets' masses and semi-major axis values, which determine the eigenmodes). In fact, the architecture of this particular system may have yielded a propensity for the outer two planets to have had their major axes aligned. Consider the normalized eigenvector components in Table 1, which represent the *relative* magnitudes of the $e$ vector for each eigenmode $m$: The absolute value is given by an amplitude factor for each eigenmode, which is determined by the initial conditions of the $e$ vectors. According to the normalized eigenvectors in Table 1, as long as those amplitude factors were comparable for all three eigenmodes, the eccentricities of both of the outer two planets would be dominated by the same mode ($m =3$), which is why there is a tendency for them to have aligned major axes.

In other words, the range of initial conditions that would have avoided setting up such an alignment of major axes would have been somewhat constrained. The eccentricity of the outer planet is about 0.35, typical for an exoplanet at that distance, and it is dominated by eigenmode $m =3$. In that case, to avoid alignment, initial conditions would have had to be such that for the middle planet the sum of the magnitudes of the $m =1$ and 2 modes is close to or greater than mode 3. Then, because for the inner planet modes 1 and 2 are each more than twice as large as for the middle planet, $e_1$ would be have been quite large. More specifically, the initial state of the system would have required $e_1$ to have been twice as large as $e_3$, or at least to get that large during its oscillations over each secular cycle. In that case, the eccentricity of the inner planet would have had to have been cycling to a value >0.7 ever since the current dynamical configuration was established. Thus, at least to the extent that second-order secular theory is indicative, the basic architecture of the system places fairly stringent constraints on the initial conditions that would have been necessary to avoid having an alignment of the major axes of the two outer planets.

Essentially, the moderate propensity for the alignment can be recognized in Table 1 by the fact that the normalized eigenvector for mode 3 (which is determined by masses and semi-major axis values only) has comparably strong effects on two planets (0.6 for planet c and 0.8 for planet d). For that reason, it is not surprising for the 61 Virginis system to show this alignment, independent of any tidal evolution. Thus, while some tidal evolution is possible, the libration about aligned pericenter longitudes is not an indication that tidal evolution played a role.

## 4.2 Other indications of eccentricity damping

On the other hand, the magnitudes of the eigenmodes are consistent with some damping of the inner planet having taken place. As described in Section 3 above, if an eccentricity-damping process acts on the inner planet, for this system mode 1 should be the fastest damping (at the rate given by $0.65F$). In fact, as shown by the absolute (not normalized) magnitude of the eigenvector in Table 1, this mode does have the lowest rms amplitude, only about 0.06 compared with 0.24 for mode 2 and 0.39 for mode 3. The damping timescale $1/(0.65F)$ is probably not less than about half the age of the system, or else the amplitude of mode 1 would be less than 0.06 by now. Damping of the other modes is consistent with that conclusion. The damping timescale for mode 2 is about twice that of mode 1, so if mode 1 had a large initial amplitude, so did mode 2. The damping rate for mode 3 is hundreds of times slower, so this process could not have changed it much.

## 4.3 Other solutions within the range of uncertainty

In this paper we have considered primarily the best-fit solution of Vogt et al. Because our main point is to demonstrate that libration of pericenters about alignment is not necessarily indicative of tidal evolution or of any other form of eccentricity damping, strictly speaking this single case is adequate to make the point.  However, it is also instructive to consider whether other sets of orbits within the range of uncertainty reported by Vogt et al. might contain signatures of tidal evolution.

In general, we find that the results are essentially unchanged for over most of the range of uncertainties of orbital elements and planetary masses.  However, the strongest indication of eccentricity damping of the inner planet appears where the mass of the outer planet is at its lower limit of 20.3 Earth masses, about 10% smaller than the best-fit (see Table 3).  With all other parameters unchanged, we find that the amplitude of the first eigenmode is now only 0.028, compared with 0.18 and 0.38 for modes 2 and 3.  Thus, the first mode could have decreased during as much as ~3 exponential-damping timescales.  In this case, the damping timescale for mode 2 is one-eighth of that for mode 1, so its amplitude is perfectly consistent with substantial eccentricity-damping having acted on the inner planet. If this solution for the mass of the third planet proves to be correct, consideration of the amplitudes of the eigenmodes suggests such damping may have occurred.

At the same time, consideration of these amplitudes shows that the libration of pericenters of the two outer planets, which is similar to that in the nominal case, is still not an indicator in itself of eccentricity damping or of a fixed-point condition.  As in the nominal case, the third eigenmode strongly affects both planets c and d, which explains the tendency to libration.  The substantial amplitude of two eigenmodes shows again that this behavior has nothing to do with the fixed-point condition.

Consider also the opposite extreme of possible values for the mass of the outer planet, the maximum value of 25.5 Earth masses (Table 4).  We find that the amplitude of mode 1, which is most closely associated with the inner planet and thus damps most quickly when the inner planet's eccentricity is damped, is 0.27.  This mode damps at the exponential rate $0.72F$ and the amplitude is not a compelling indicator of much eccentricity damping. At most it has undergone less than one damping timescale over the age of the system. Interestingly, one mode (labeled #2 here) has a low amplitude of only 0.12, which might be suggestive of damping except that it should damp more slowly than mode 1, which has not damped very much. Here again the pericenters of the outer two plants librate about one another, but only because of their intrinsic eigenvector for mode 3. This case is an even better demonstration of libration independent of tidal evolution than the nominal case is.

In all our calculations so far we have simply assigned the value of the minimum mass ($M$ sin $i$) to the mass of each planet. Of course it is most likely that the actual masses are somewhat larger. If we raise the values of all the masses in proportion to one another, the effect is similar to increasing the mass of the outer planet alone, as described above. Tables 5 and 6 show the results for a 10% and 20% increase, respectively, in the masses relative to the observational minimum mass. The results for these cases are similar to those for increasing $m_3$ alone. Again we see that, although one mode has an amplitude small enough to suggest eccentricity damping, the mode that should damp fastest still has a substantial amplitude. So again, for either of these cases, there is no evidence for tidal damping. And again the pericenters of the outer two planets are locked into mutual libration, independent of any tidal evolution.

As discussed in Section 4.1 above, the tendency for libration, rather than circulation, of the pericenters of the outer two planets is a consequence of the (normalized) eigenvector for mode 3 having fairly large components for both of those planets. The eigenvector depends on the masses and semimajor axis values in the system. Nevertheless it is possible to have values of e and $\varpi$ values at some given time that yield amplitudes $E$ for the other modes that can overcome this tendency. For example, if we assume that the system is described by Vogt et al.'s nominal parameters, but change $\varpi_1$ and $\varpi_2$ to the maximum and minimum values, respectively, we do find that the system circulates. This result is evident in Table 7, where we see that for planet 2 the magnitude of the eccentricity vector components $E_1V_{12}$ and $E_2V_{22}$ combined are greater than $E_3V_{32}$, (i.e. for this planet modes 1 and 2 combined are stronger than mode 3). Hence the pericenter longitude of planet 2 is not bound to that of planet 3. In this case there is no libration, but the issue is completely independent of whether or not there has been tidal evolution.

## 4.4 The fixed-point condition and the inclination degeneracy

In general solutions for the orbital elements of extra-solar planets from observations suffer from an ambiguity in the inclination $I$ of an orbit relative to the line of sight. The value of $I$ cannot be uncoupled from the masss of the planet. However, in principle, if a multi-planet system can be reasonably assumed to have been damped to a state where only a single eigenmode remains with significant amplitude, the number of unknown parameters is reduced, and it may be possible to solve for $I$ along with other orbital elements (Batygin and Laughlin 2011).

The practical difficulty, of course, is that it may not be possible to determine *a priori* whether a system is in such a fixed-point condition. Batygin and Laughlin (2011) considered the 61 Virginis system as an example of a system where such a condition might be expected, especially if one assumed a small value of the tidal parameter Q, typical of a terrestrial planet, for the inner planet. However, as we have shown here, the range of orbital solutions for 61 Virginis seems to preclude the liklihood that this particular system is in a fixed-point condition. It is noteworthy however that for this system the slowest damping eigenmode is very long-lived compared with the other eigenmodes (assuming most dissipation is in the inner planet; e.g. Table 2), which demonstrates that once a fixed-point condition is reached, it may endure for a large part of the life of the system. If typical, this result suggests that fixed-point conditions may be common enough that solutions for $I$ may indeed be possible in a significant number of cases.

## 5. Conclusion

In any planetary system where orbital behavior is dominated by secular interactions, an alignment of the major axes, or libration about such an alignment, would occur if one eigenfrequency were dominant. Such dominance could result (a) from damping of all but the longest-lived eigenmode, or (b) from exponentially increasing the amplitude of one eigenmode relative to the others in some way (Chiang and Murray 2002). In the nominal 61 Virginis case, no single eigenmode is dominant, so there is no evidence for evolution of type (a) or (b). Most likely the current libration about alignment of the outer two planets' orbits is due to fortuitous initial conditions. Moreover, in this system the intrinsic strength of the $m=3$ eigenmode for those two planets (a consequence of system's architecture) contributes to a moderate propensity for such alignment.

For the 61 Virginis system the libration of the outer two planets appears to be a consequence of the basic architecture of this planetary system and of the inner planet having formed with a moderate eccentricity. There is no evidence of tidal evolution. These conclusions would remain essentially unchanged for any values of the masses and semi-major axes within the range of observational uncertainty shown in Table 1. Even if subsequent observations show the parameters of this system to be quite different from reported values (which is quite possible), the basic dynamical principle discussed here may be relevant more generally: An observed alignment of some of the major axes in a planetary system does not imply that one eigenmode is dominant, and thus cannot in itself suggest that tidal evolution or eccentricity damping by any other mechanism has taken place.

Alignment of major axes (or libration about alignment) is not in itself a valid test for possible cases of tidal evolution. Classical secular theory does show that such alignment can be a result of eccentricity damping, but it could also result from the basic architecture of the system combined with appropriate initial conditions. Instead, a better test is consideration of the amplitudes of the eigenmodes. Unless the modes that should damp fastest actually have small amplitudes, there is no evidence for eccentricity damping in a multi-planet system.

Finally we emphasize that this discussion has been in the context of classical analytic secular theory. Indeed, this theory was the implicit basis for the discussion by Vogt et al. of the possibility of tidal evolution in the 61 Virginis system. Of course this theory requires that the eccentricities are not so great that the second-order theory is invalid, a limit whose boundaries remain poorly defined. Nevertheless, classical secular theory can be a powerful tool for characterizing the dynamical state of planetary systems in ways that can guide consideration of their origin and evolution.

**Appendix A: Secular theory, notation, and definitions**

Define elements $h_p = e_p \sin \varpi_p$ and $k_p = e_p \cos \varpi_p$, where integer $p$ identifies the planets in order of increasing semi-major axis. Thus the eccentricity vector (magnitude $e_i$ and direction $\varpi_i$) is

$$\vec{e_p} \equiv \left( k_p, h_p \right) \qquad\qquad \text{Eq. (A1)}$$

Following classical secular theory (e.g. Brouwer and Clemence 1961, Murray and Dermott, 1999), retaining terms in the disturbing potential to second-order in eccentricities, ignoring terms that contain orbital longitudes, and assuming that the orbits are coplanar, the planet's mutual perturbations yield:

$$\dot{h}_p = \sum A_{pj} k_j \quad \text{and} \quad \dot{k}_p = -\sum A_{pj} h_j \qquad\qquad \text{Eq. (A2)}$$

where the coefficients $A_{pj}$ are well-known functions of masses and semi-major axes. Following Vogt et al. (2010), we include a contribution to the precession from general relativity, by adding a term $3 n_p^3 a_p^2 / c^2$ in $A_{pp}$. The linear differential equations have solutions of the form:

$$k_p = K_p e^{i(gt + \delta)} \text{ and } h_i = -k_i i \qquad\qquad \text{Eq. (A3)}$$

where $i \equiv (-1)^{1/2}$. Substituting (A3) into either part of (A2) yields the same equation:

$$K_p g = \sum A_{pj} K_j \qquad\qquad \text{Eq. (A4)}$$

where $(K_1, K_2, K_3)$ is the eigenvector corresponding to the eigenfrequency $g$.

Here, with three planets, we can solve the three equations (A5) for the three unknowns $\omega$, $x_1 \equiv K_1/K_3$, $x_2 \equiv K_2/K_3$. (The solution gives ratios of the components of each eigenvector, here normalized to $K_3$.) There are three solutions for $g$, which we call $g_m$ (with m =1 to 3), and each of those three eigenfrequencies has a corresponding eigenvector $(K_{m1}, K_{m2}, K_{m3})$. Note that in the latter expression, the subscripts contain the mode $m$ followed by a number identifying a planet.

Standard analytical or numerical solutions of (A4) yield the values of the three eigenfrequencies $g_m$ and the corresponding values of $x_1$ and $x_2$ for each eigenmode.

Because for each eigenmode we know the ratios of the $K$ values, it is convenient to express them in terms of the rms of their values, $E_m$, so that $K_{mp} = E_m V_{mp}$. By definition, for each eigenmode the ratios of the $V_{mp}$ values are the same (known) ratios of the corresponding $K_{mp}$ values, and the rms of $V_{m1}$, $V_{m2}$, and $V_{m3}$ is 1. Thus the value of $V_{mp}$ is known from the above solution for values of $m$ and $p$.

The solution for the behavior of elements $k_p$ and $h_p$ for each planet $p$ now takes the form (c.f. A3):

$$k_p = \sum E_m V_{mp} \cos(g_m t + \delta_m)$$     Eq. (A5a)

$$h_p = \sum E_m V_{mp} \sin(g_m t + \delta_m)$$     Eq. (A5b)

The solution includes six constants of integration, two corresponding to each of the three eigenmodes: the phase $\delta_m$ and the magnitude of the eigenvector $E_m$. The six constants of integration are determined by inserting the known values of $k_p$ and $h_p$ (equivalently $e_p$ and $\varpi_p$) for all three planets into equations (A5) and solving those six equations for the six unknowns $\delta_m$ and $E_m$.

According to Eqs. (A5), for each planet $i$ the eccentricity vector $\mathbf{e}_p \equiv (k_p, h_p)$ is the sum of three vectors $\mathbf{e}_{mp}$ (corresponding to the three modes $m$), each with a constant length $E_m V_{mp}$, that rotate at independent rates given by their corresponding eigenfrequencies $g_m$ (Fig. 1). Also, for any eigenmode, the eccentricity vectors for the three planets ($\mathbf{e}_{m1}$, $\mathbf{e}_{m2}$ and $\mathbf{e}_{m3}$) must be parallel or anti-parallel (depending on the signs of the $V_{mp}$ values).

**Appendix B: Effect of eccentricity damping on secular behavior**

Suppose some process acts to damp the eccentricity of the inner planet according to

$$\mathrm{d}e_1/\mathrm{d}t = -F e_1$$     Eq. (B1)

where $F$ can be considered nearly constant. This effect can be incorporated into the secular theory by adding a term $-Fh_1$ to $dh_1/dt$ and $-Fk_1$ to $dk_1/dt$ in Eq. (A2). The secular equations are still linear, so solutions will still be of the form (A3). Thus Eq. (A4) becomes (in matrix form):

$$[\mathbf{B}]\mathbf{K} - g\mathbf{K} = 0 \qquad\qquad \text{Eq. (B2)}$$

where the matrix $[\mathbf{B}]$ replaces $[\mathbf{A}]$ from Eq. (A4) with $B_{11} = A_{11} + Fi$, and all other $B_{ij} = A_{ij}$.

Solution of these linear equations for $g$, $x_1$, and $x_2$ will again yield three sets of solutions (the eigenmodes m = 1 to 3), now with complex eigenfrequencies whose imaginary parts yield damping rates for the eigenvectors when inserted into (A3). There are various ways to solve the equations. Here we take advantage of the fact that $F$ is relatively small because the damping is weak, so we can use a perturbation approach based on the solution with $F$=0 (Appendix A). Recalling that the solution to the case with $F$=0 is $(g, x_1, x_2)$, for small $F$, we express the solution set as $(g + \varepsilon, x_1 + \varepsilon_1, x_2 + \varepsilon_2)$ where $\varepsilon$, $\varepsilon_1$, and $\varepsilon_2$ must be small. Inserting these expressions into Eq. (B2) yields

$$\left[\mathbf{B}\right]\begin{bmatrix} x_1 + \varepsilon_1 \\ x_2 + \varepsilon_2 \\ 1 \end{bmatrix} - (g + \varepsilon)\begin{bmatrix} x_1 + \varepsilon_1 \\ x_2 + \varepsilon_2 \\ 1 \end{bmatrix} = 0 \qquad\qquad \text{Eq. (B3)}$$

Expanding these three equations, retaining only first order terms in $\varepsilon$, $\varepsilon_1$, and $\varepsilon_2$, and subtracting out the corresponding unperturbed equations (A4), we obtain

$A_{11}\varepsilon_1 + A_{32}\varepsilon_2 = \varepsilon$

$q_1\varepsilon_1 + q_2\varepsilon_2 = -x_1 Fi$ \qquad\qquad Eqs. (B4)

$q_3\varepsilon_1 + q_4\varepsilon_2 = 0$

where $q_1 \equiv A_{11} - g - A_{31}x_1$, $q_2 \equiv A_{12} - A_{32}x_1$, $q_3 \equiv A_{21} - A_{31}x_2$, and $q_4 \equiv A_{22} - g - A_{32}x_2$.

Eqs. (B4) are linear in $\varepsilon$, $\varepsilon_1$, and $\varepsilon_2$, and can be readily solved in closed form, yielding $\varepsilon$, whose imaginary part below is the damping rate corresponding to each eigenfrequency:

$\varepsilon = [(A_{32}q_3 - A_{31}q_4)/(q_1q_4 - q_2q_3)]x_1 Fi$ . \qquad\qquad Eq. (B5)

**Acknowledgments**

This work was supported by a grant from NASA's Planetary Geology and Geophysics program. Christa Van Laerhoven's participation was also supported by a Canadian NSERC Postgraduate Scholarship.

Figures:

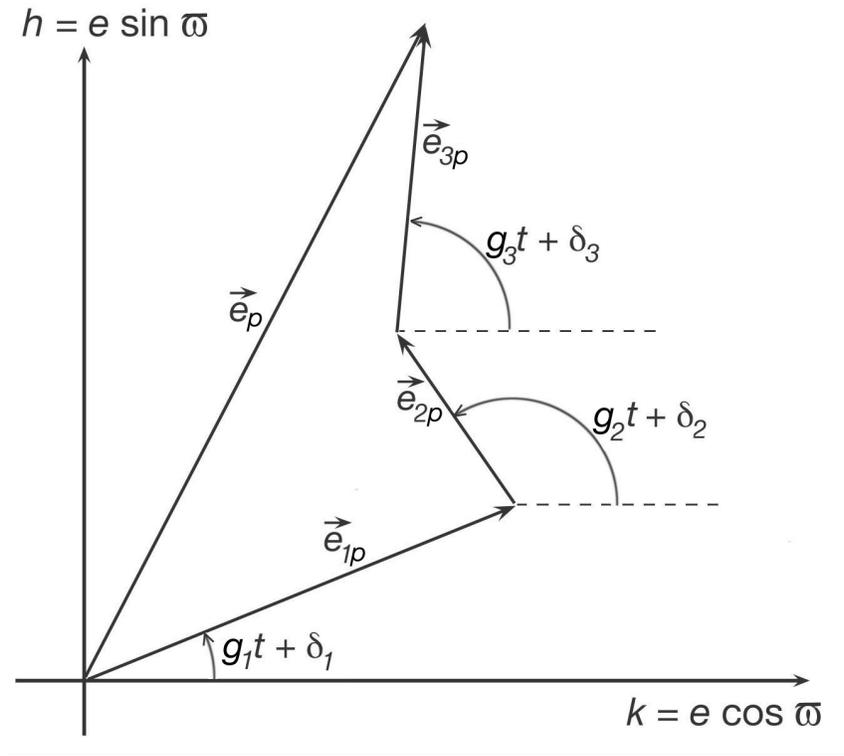

<u>Figure 1</u>:  Behavior of the eccentricity vector $\mathbf{e}_p \equiv \left(k_p, h_p\right)$ of planet $p$, according to the secular solution (Appendix A).  The magnitude of each component vector $\mathbf{e}_{mp}$ is $E_m V_{mp}$ and its direction rotates at the rate given by the eigenfrequency $g_m$, with phase $\delta_m$ given by initial conditions.

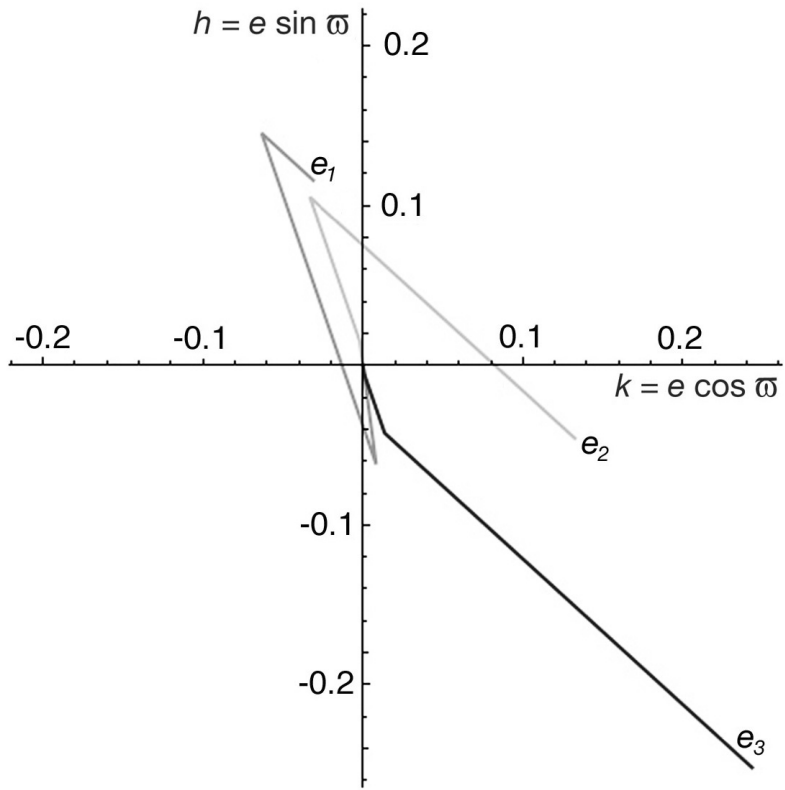

Figure 2: The eccentricity vectors $(k,h)$ for planets 1, 2, and 3 (that is 61 Vir b, c, and d, shown in gray, lighter gray, and black, respectively), each shown as a vector sum of its three eigenfrequency components. In this diagram, for each planet, the component corresponding to eigenmode $m = 1$ starts at the origin, the one for $m = 2$ starts at the tip of the first component, and the one for $m = 3$ starts at the tip of that component and terminates at the total magnitude and direction of the eccentricity vector of that planet.

**Table 1:** Orbital elements and masses for the 61 Virginis system from Vogt et al. (2010).

| Planet | Period (days) | $e$ | $\varpi$ | $M \sin i$ (*Earth masses*) | $a$ (AU) |
|---|---|---|---|---|---|
| 61 Vir | | | | | |
| b | 4.2150 ± 0.0006 | 0.12 ± 0.11 | 105° ± 54° | 5.1 ± 0.5 | 0.0502 ± 5×10⁻⁶ |
| c | 38.021 ± 0.034 | 0.14 ± 0.06 | 341° ± 38° | 18.2 ± 1.1 | 0.2175 ± 0.0001 |
| d | 123.01 ± 0.55 | 0.35 ± 0.09 | 314° ± 20° | 22.9 ± 2.6 | 0.476 ± 0.001 |

**Table 2:** Eigenmodes for the secular solution assuming the nominal (best-fit) elements and masses from Table 1. The normalized eigenvectors $V_{mp}$ (normalized to the rms value for each eigenmode) and the eigenfrequency $g_m$ are based only on masses and semi-major axes. The amplitudes of each eigenmode $E_m$, are constants of integration determined from values of eccentricities and pericenter longitudes at the time of observation. Shown in parentheses are the actual magnitudes of each eigenvector component, given by $E_m V_{mp}$. The last two columns show the damping rates for the eigenmodes: #1 is due to a damping process acting on planet 1, as developed in Appendix B, and on planet 2.

| Eigen-mode $m$ | Eigenvector $V_{mp}$ (and $E_m V_{mp}$) | | | $g_m$ (°/yr) | $E_m$ | Damping Rate $(\mathrm{d}E_m/\mathrm{d}t)/E_m$ | |
|---|---|---|---|---|---|---|---|
| | Planet 1 "b" | Planet 2 "c" | Planet 3 "d" | | | Planet 1 damped | Planet 2 damped |
| 1 | 0.968 (0.062) | -0.236 (-0.015) | 0.081 (0.005) | 0.039 | 0.064 | -0.65F | -0.29$F_2$ |
| 2 | 0.904 (0.219) | 0.396 (0.096) | -0.163 (-0.039) | 0.035 | 0.24 | -0.35F | -0.50$F_2$ |
| 3 | 0.114 (0.044) | 0.581 (0.224) | 0.806 (0.311) | 0.0092 | 0.39 | -0.0011F | -0.22$F_2$ |

**Table 3:** Similar to Table 2, except that here the outer planet's mass is taken to be 20.3 Earth masses, the minimum value of $M \sin i$ from Table 1.

| Eigen-mode $m$ | Eigenvector $V_{mp}$ (and $E_m V_{mp}$) | | | $g_m$ (°/yr) | $E_m$ | Damping Rate $(dE_m/dt)/E_m$ | |
|---|---|---|---|---|---|---|---|
| | Planet 1 "b" | Planet 2 "c" | Planet 3 "d" | | | Planet 1 damped | Planet 2 damped |
| 1 | 0.992 (0.027) | -0.123 (-0.003) | 0.044 (0.001) | 0.038 | 0.028 | -0.88$F$ | -0.10$F_2$ |
| 2 | 0.727 (0.133) | 0.623 (0.114) | -0.289 (-0.053) | 0.033 | 0.183 | -0.12$F$ | -0.65$F_2$ |
| 3 | 0.115 (0.044) | 0.592 (0.226) | 0.797 (0.304) | 0.0090 | 0.381 | -0.0013$F$ | -0.25$F_2$ |

**Table 4:** Similar to Table 2, except that here the outer planet's mass is taken to be 25.5 Earth masses, the maximum value of $M \sin i$ from Table 1.

| Eigen-mode $m$ | Eigenvector $V_{mp}$ (and $E_m V_{mp}$) | | | $g_m$ (°/yr) | $E_m$ | Damping Rate $(dE_m/dt)/E_m$ | |
|---|---|---|---|---|---|---|---|
| | Planet 1 "b" | Planet 2 "c" | Planet 3 "d" | | | Planet 1 damped | Planet 2 damped |
| 1 | 0.979 (0.264) | 0.192 (0.052) | -0.074 (-0.020) | 0.037 | 0.270 | -0.73$F$ | -0.21$F_2$ |
| 2 | 0.867 (0.107) | -0.474 (-0.058) | 0.153 (0.019) | 0.041 | 0.123 | -0.27$F$ | -0.60$F_2$ |
| 3 | 0.113 (0.044) | 0.572 (0.223) | 0.812 (0.316) | 0.0093 | 0.389 | -0.001$F$ | -0.19$F_2$ |

**Table 5:** Similar to Table 2, except that here all three planets' masses are taken to be 10% larger than the nominal values of $M \sin i$ from Table 1, *i.e.* the orbit normals do not lie in the plane of the sky.

| Eigen-mode $m$ | Eigenvector $V_{mp}$ (and $E_m V_{mp}$) | | | $g_m$ (°/yr) | $E_m$ | Damping Rate ($\mathrm{d}E_m/\mathrm{d}t)/E_m$ | |
|---|---|---|---|---|---|---|---|
| | Planet 1 "b" | Planet 2 "c" | Planet 3 "d" | | | Planet 1 damped | Planet 2 damped |
| 1 | 0.958 | 0.262 | -0.112 | 0.038 | 0.274 | -0.57$F$ | -0.32$F_2$ |
| 2 | 0.926 | -0.355 | 0.128 | 0.042 | 0.110 | -0.43$F$ | -0.46$F_2$ |
| 3 | 0.120 | 0.581 | 0.805 | 0.010 | 0.386 | -0.001$F$ | -0.22$F_2$ |

**Table 6:** Similar to Table 5, except that here all three planets' masses are taken to be 20% larger than the nominal values of $M \sin i$ from Table 1.

| Eigen-mode $m$ | Eigenvector $V_{mp}$ (and $E_m V_{mp}$) | | | $g_m$ (°/yr) | $E_m$ | Damping Rate ($\mathrm{d}E_m/\mathrm{d}t)/E_m$ | |
|---|---|---|---|---|---|---|---|
| | Planet 1 "b" | Planet 2 "c" | Planet 3 "d" | | | Planet 1 damped | Planet 2 damped |
| 1 | 0.980 | 0.183 | -0.083 | 0.041 | 0.277 | -0.73$F$ | -0.19$F_2$ |
| 2 | 0.863 | -0.475 | 0.175 | 0.045 | 0.127 | -0.26$F$ | -0.59$F_2$ |
| 3 | 0.127 | 0.581 | 0.804 | 0.011 | 0.386 | -0.001$F$ | -0.22$F_2$ |

**Table 7:** Similar to Table 2, except that here $\varpi_2$ has been increased to the maximum value 379° and $\varpi_3$ has been decreased to the minimum value 294°.

| Eigen-mode $m$ | Eigenvector $V_{mp}$ (and $E_m V_{mp}$) | | | $g_m$ (°/yr) | $E_m$ | Damping Rate ($\mathrm{d}E_m/\mathrm{d}t)/E_m$ | |
|---|---|---|---|---|---|---|---|
| | Planet 1 "b" | Planet 2 "c" | Planet 3 "d" | | | Planet 1 damped | Planet 2 damped |
| 1 | 0.968 (0.230) | -0.236 (-0.056) | 0.081 (0.019) | 0.039 | 0.237 | -0.65$F$ | -0.29$F_2$ |
| 2 | 0.904 (0.369) | 0.396 (0.161) | -0.163 (-0.066) | 0.035 | 0.408 | -0.35$F$ | -0.50$F_2$ |
| 3 | 0.114 (0.039) | 0.581 (0.201) | 0.806 (0.279) | 0.0092 | 0.346 | -0.0011$F$ | -0.22$F_2$ |